\documentstyle[psfig]{espart}
\begin{document}
\begin{frontmatter}
  \title{{\normalsize J. Stat. Phys. Vol.84, 1373 (1996)}\\[1.5cm]
Avalanche statistics of sand heaps} \author{Volkhard
    Buchholtz and Thorsten P\"oschel} \address{Humboldt-Universit\"at
    zu Berlin, Institut f\"ur Physik, Invalidenstra\ss e 110, D-10115
    Berlin, Germany, volkhard@itp02.physik.hu-berlin.de,
    thorsten@itp02.physik.hu-berlin.de\\ URL:
    http://summa.physik.hu-berlin.de/}

\maketitle
\begin{abstract}
  Large scale computer simulations are presented to investigate the
  avalanche statistics of sand piles using molecular dynamics. We
  could show that different methods of measurement lead to
  contradicting conclusions, presumably due to avalanches not reaching
  the end of the experimental table.
\end{abstract}
\end{frontmatter}
The physics of an evolving sandpile has been of large interest to
physicists and engineers and there has been done much work in this
field. One of the most popular (or sometimes unpopular) ideas is the
concept of self organized
criticality (SOC)~\cite{BakTangWiesenfeld:1987}.
It has been argued by many physicists that sandpiles can be described
by cellular automata in two or three dimensions
(e.g.~\cite{WiesenfeldTangBak:1989})
and by stochastic cellular automata (e.g.~\cite{Frette:1993}) which
in simulations might show SOC-behavior. There are many effects in nature
which are supposed to reveal SOC, and hence there is a variety of
articles investigating the theory of
SOC~(e.g.~\cite{KerteszKiss:1990ug}).
When particles are dropped one after the other onto the top of a sand
heap one observes avalanches. The time intervals in between  successive
avalanches and the size distribution of the avalanches have been of
large interests to many scientists. Experimentalists as well as
theorists investigated these quantities, and there is a controversy whether
they obey a power law or
not~\cite{KadanoffNagelWuZhou:1989,RosendahlVekicKelley:1993}.

In an experimental work Jaeger et al.~\cite{JaegerLiuNagel:1989}
investigated the avalanche sizes of a pile contained in a box with one
open side. The material flow over the edge of the box was measured
in between a pair of capacitor plates. From the fluctuation
of the capacity they concluded that the sizes of the avalanches might {\em not} be
power law distributed. They count for the size of an avalanche the
capacity change, i.e. the mass of the particles which fall down over
the edge of the table. Bretz et al. measured the avalanche
distribution by recording the temporal behavior of the inclination of
the heap's surface~\cite{BretzCunninghamKurczynskiNori:1992}.
Contradicting to~\cite{JaegerLiuNagel:1989} their results support the
hypothesis of the power law distribution.

We want to present the results of a large scale computer experiment where we
recorded the distribution of the sizes of avalanches using both
methods. Finally we will show that, perhaps, both
measurements~\cite{JaegerLiuNagel:1989,BretzCunninghamKurczynskiNori:1992}
do not contradict, but support each other.

\begin{figure}[htbp]
  \centerline{\psfig{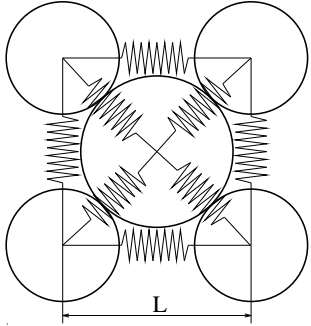}}
  \caption{Each of the non spherical particles consists of five spheres.}
  \label{AmO}
\end{figure}

In a previous paper~\cite{PoeschelBuchholtz:1993} we have shown that
in simulation using two dimensional molecular dynamics of non
spherical particles one can find a power law behavior of the size
distribution of the avalanches. Our particles $k$ have been built up
of five spheres which are connected by springs~(fig.~\ref{AmO}).

At rest the inner sphere touches the surrounding spheres of the same
particles. For details of the forces acting between the particles of
the same grain via springs and the forces acting between colliding
grains see~\cite{BuchholtzPoeschel:1994PA}.

Using molecular dynamics we build up the pile by dropping the
particles one after the other on the top of the evolving pile. A
particle was dropped when all avalanches and fluctuations caused by
the previously released grain have faded away, i.e. we waited
until the maximum velocity of the particles comes very close to zero.
Then the inclination of the heap was measured due to the following
procedure: Suppose the shape of the heap of height $H$ built up on a
surface of width $B$ is close to a triangle. Then its slope is given
by
\begin{equation}
  \label{slope}
  \Phi =
  \mbox{arctan}\left[\frac{H-\frac{1}{M}\sum\limits_{i=1}^{N}m^{(k)}y^{\,(k)}}{\frac{2}{M}\sum\limits_{i=1}^{N}m^{(k)}x^{\,(k)}}\right]~,
\end{equation}
where $m^{(k)}$, $x^{\,(k)}$ and $y^{\,(k)}$
the mass and the position of the $k$th grain and $M$ is the sum of the masses of all particles $M=\sum_{i=1}^N m^{(k)}$.
Since our heap is close to, but not an ideal triangle we calculated
the height $H$ using
\begin{equation}
  \label{hight}
  H=\frac{2}{x_{max}} \int_0^{x_{max}} h(x) dx~,
\end{equation}
where $x_{max}$ is the $x$-position of the grain which is closest to
the end of the table.  From the fluctuations of the slope due to
eq.~(\ref{slope}) we can conclude the approximate size of the
avalanche according to a decrease in the slope:
\begin{equation}
  \label{fluct1}
  \Delta^{(1)} M = \frac{B^2}{2}\cdot\rho (\tan\Phi' - \tan\Phi),
\end{equation}
where $\Phi'$ is the slope before the dropping event, $B\approx P=30.7
cm$ is the length of the table and $\rho = 0.59~ g\cdot cm^{-2}$
denotes the average density of the heap. This method for the
measurement of the size of an avalanche (indexed by $\Delta^{(1)} M$)
is close to the experimental method used by Bretz et
al.~\cite{BretzCunninghamKurczynskiNori:1992}. Another method which
was used by Jaeger et al.~\cite{JaegerLiuNagel:1989} and by Rosendahl
et al.~\cite{RosendahlVekicKelley:1993} is to measure the weight of
the material which reaches the end of the finite table during an
avalanche using a balance or a capacitor. They considered the weight
of material flowing over the border of the table to be the size of the
avalanche $\Delta^{(2)} M$. Similar as in the experiment in the
molecular dynamics simulation we calculated the mass of the particles
which reach the end of the table, i.e. $x^{(k)}>P$.

Collecting the idle time of all computers of our department over a
period of about one year we found enough computer power to perform a
large scale molecular dynamics simulation. The heap was build up on a
rough surface of width $P=30.7~cm\approx 200 \cdot
\left<{r_i^{(k)}}\right>$. The radii $r_i^{(k)}$ of the outer spheres of
the particles $k$ were equally distributed in the interval
$r_i^{(k)}\in (0.05,0.11)~cm$ while the radii of the inner spheres are
determined by the relation $r_m^{(k)}= L^{(k)}/\sqrt{2} - r_i^{(k)}$
($i=1\dots 4$). $L^{(k)}$ is the size of the $k$th grain (s. fig.~\ref{AmO}). In~\cite{BuchholtzPoeschel:1994PA} (fig.~10) we have
shown that the relation $\frac{L^{(k)}}{r_i^{(k)}} = 4$ reproduces
well the static friction behavior of sandpiles. The average number of
particles on the heap was $N_{av}=930$.

\begin{figure}[htbp]
  \centerline{\psfig{figure=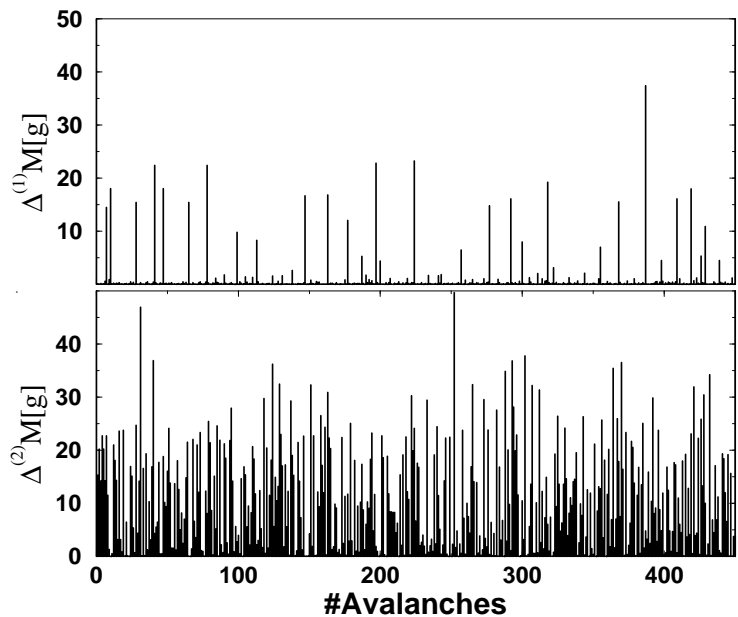,width=10cm,angle=270}}
  \caption{Series of 450 avalanches. Top: the avalanche 
    size $\Delta^{(1)} M$ concluded from the fluctuation in slope
    (equation~(3)), 
bottom: the avalanche size
    $\Delta^{(2)} M$ calculated from the mass of particles that reach
    the end of the table. The fraction of small avalanches is much
    higher for the upper figure.}
  \label{fig:TimeSeries}
\end{figure}
\begin{figure}[htbp]
  \centerline{\psfig{figure=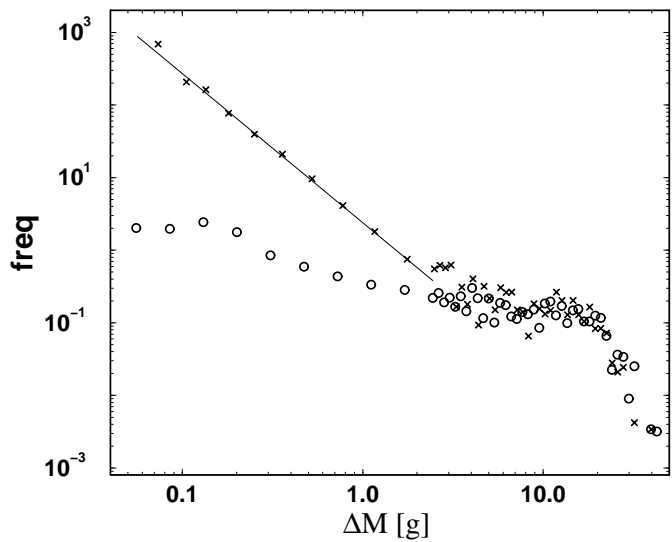,width=10cm,angle=270}}
  \caption{The size distribution of the avalanches in log-log-scale due to 
    $\Delta^{(1)} M$ ($\diamond$) and $\Delta^{(2)} M$ ($+$),
    respectively. The sizes are measured in gramms. The line shows
    the function $freq\sim \Delta^{(1)} M^{-1.85}$}
  \label{fig:SizeDistribution}
\end{figure}

Figure~\ref{fig:TimeSeries} shows the time series of the avalanche size
due to both procedures, $\Delta^{(1)} M$ and $\Delta^{(2)} M$.
The size distributions of the data shown in fig.~\ref{fig:TimeSeries}
are drawn in fig.~\ref{fig:SizeDistribution}.
We find that the distribution of the avalanche sizes measured by means
of the fluctuations in the slope $\Delta^{(1)} M$ reveals a typical
power law behavior for avalanches smaller than $\Delta M^{(1)} <
2~\mbox{gramms}$, while the distribution according to the direct measurement of
the avalanche sizes $\Delta^{(2)} M$ does {\em not} show a power
law behavior. We claim that the difference in both figures
(\ref{fig:SizeDistribution}) comes from the fact that the second
method is not able to care for those avalanches which do not reach the
end of the table. Obviously the larger the pile the higher is the
fraction of avalanches which do not reach the border of the table.
This coincides with the observations by Jaeger et
al.~\cite{JaegerLiuNagel:1989} who found a deviation from the power
law scaling for {\em large} systems. They found a sharply peaked
avalanche distribution of large, system overspanning avalanches.

For large avalanches in our simulation both types of measurements lead
to very similar results which supports our conclusion. In agreement
with the experimental observations by Rosendahl et
al.~\cite{RosendahlVekicKelley:1993} we find large avalanche tails in
the distribution. In the case of large piles, the direct measurement
of the mass fluctuations, i.e. neglecting the smaller avalanches,
would lead to similar results as Jaeger et.~al. found.

\begin{figure}[htbp]
   \centerline{\psfig{figure=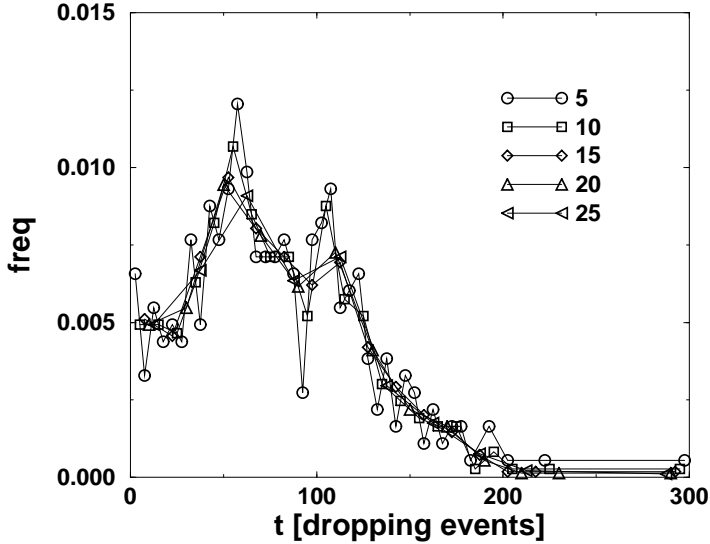,width=10cm,angle=270}}
  \caption{The waiting time distribution of the intervals in between
    two consecutive large avalanches. The figure shows the number of
    pairs of avalanches over their time distance measured in dropping
    events. The double peak structure remains preserved for different
    discretization interval sizes.}
  \label{Poisson} 
\end{figure}

For the waiting time distribution of the large avalanches, i.e. for
the distribution of the number of dropping events in between two
consecutive large avalanches, surprisingly we find a double peak.
Fig.~\ref{Poisson} shows the distribution for five different sizes of
the discretization intervals. The double peak structure is found in
all five curves, hence we assume that it is not an artifact due to
the choice of the size of the discretization interval. So far we have
no explanation for this behavior.

Although our simulation seems at least not to contradict the concept
of SOC we should remark here that there are other serious objections
against applying the idea of SOC in the case of sand pile avalanches
(see e.g.~\cite{Evesque:1991}).

\ack 
We thank the members of the institute for their patience due to
inconveniences connected with the permanently on all computers running
job ``HaufenGross''. T.~P. thanks H.~Jaeger, H.~J.~Herrmann and W.~Ebeling for
discussion. V.~B. has been supported by Deutsche
Forschungsgemeinschaft (RO 548/5-1).

\end{document}